\documentclass[twocolumn,tighten,times]{aastex631}
\usepackage{color}
\usepackage{multirow}
\usepackage{chngcntr}
\usepackage{mathtools}
\usepackage[utf8x]{inputenc}
\usepackage{perpage}
\MakePerPage{footnote}
\usepackage{lipsum}
\usepackage[graphicx]{realboxes}
\shorttitle{UDGs with bulges}
\shortauthors{Rong et al.}

\begin{document}

\title{Ultra Diffuse Dwarf Galaxies Hosting Pseudo-bulges}

\correspondingauthor{Yu Rong}
\email{rongyua@ustc.edu.cn}

\author{Yu Rong}
\affiliation{Department of Astronomy, University of Science and Technology of China, Hefei, Anhui 230026, China}
\affiliation{School of Astronomy and Space Sciences, University of Science and Technology of China, Hefei 230026, Anhui, China}

\author{Hong-Xin Zhang}
\affiliation{Department of Astronomy, University of Science and Technology of China, Hefei, Anhui 230026, China}
\affiliation{School of Astronomy and Space Sciences, University of Science and Technology of China, Hefei 230026, Anhui, China}

\author{Cheng Cheng}
\affiliation{National Astronomical Observatories, Chinese Academy of Sciences, Beijing 100012, China}

\author{Qi Guo}
\affiliation{National Astronomical Observatories, Chinese Academy of Sciences, Beijing 100012, China}

\author{Weiyu Ding}
\affiliation{Department of Astronomy, University of Science and Technology of China, Hefei, Anhui 230026, China}
\affiliation{School of Astronomy and Space Sciences, University of Science and Technology of China, Hefei 230026, Anhui, China}

\author{Zichen Hua}
\affiliation{Department of Astronomy, University of Science and Technology of China, Hefei, Anhui 230026, China}
\affiliation{School of Astronomy and Space Sciences, University of Science and Technology of China, Hefei 230026, Anhui, China}



\author{Huiyuan Wang}
\affiliation{Department of Astronomy, University of Science and Technology of China, Hefei, Anhui 230026, China}
\affiliation{School of Astronomy and Space Sciences, University of Science and Technology of China, Hefei 230026, Anhui, China}

\author{Xu Kong}
\affiliation{Department of Astronomy, University of Science and Technology of China, Hefei, Anhui 230026, China}
\affiliation{School of Astronomy and Space Sciences, University of Science and Technology of China, Hefei 230026, Anhui, China}



\begin{abstract}

	By analyzing data from DESI Legacy Imaging Survey of the dwarf galaxies in the Arecibo Legacy Fast Alfa Survey, we have identified five ultra-diffuse galaxies (UDGs) featuring central pseudo-bulges. These UDGs display blue pseudo-bulges with S\'ersic indices $n<2.5$ and effective radii spanning 300-700 pc, along with bluer thin stellar disks exhibiting low surface brightness and expansive effective radii that align with the UDG definition. The rotation velocities of these UDGs, determined using HI line widths and optical inclinations, exceed those of most dwarf galaxies of similar mass, suggesting the high halo spins or substantial dark matter halos. We propose that these UDGs likely formed through mergers of dwarf galaxies lacking old stars in their progenitors, resulting in the development of central bulge-like structures during starbursts triggered by the mergers, while also enhancing their halo spin. Subsequent gas accretion facilitated the formation of extended stellar disks. It is also worth noting the possibility that these UDGs could alternatively represent  ``failed $L^{\star}$ galaxies'' with massive dark matter halos but reduced star formation efficiencies. If future high-resolution HI observations confirm the presence of massive halos around these UDGs, they may have formed due to intense AGN feedback in the early universe, and may be the descendants of ``little red dots'' observed by the James Webb Space Telescope, which are characterized by heightened central black hole masses and intensified accretion and feedback processes in the early universe.

\end{abstract}


\section{Introduction} \label{sec:1}

Ultra-diffuse galaxies \citep[UDG;][]{vanDokkum15}, characterized by effective radii comparable to that of the Milky Way but with 100-1,000 times fewer stars, have garnered significant attention due to their unique properties and elusive formation mechanisms. UDGs are ubiquitously distributed throughout galaxy clusters, groups, and low-density environments. Similar to conventional dwarf galaxies, a substantial number of UDGs residing in galaxy clusters and groups present nuclear star clusters \citep[NSC; e.g.,][]{Yagi16,Lambert24,Marleau21,Rong19}. It has been postulated that certain ultra compact dwarfs (UCDs) may be the remnants of nucleated UDGs after tidal stripping \citep{Mihos17,Janssens17,Janssens19}.

The formation mechanism of UDGs remains enigmatic. Cosmological simulations offer insights into a high-spin UDG formation scenario \citep{Amorisco16,Rong17a,Liao19,Benavides23}, indicating that UDGs may acquire heightened specific angular momenta from their high-spin halos, a hypothesis supported by certain observational evidence \citep[e.g.,][]{Rong24a,Rong20c}. However, zoom-in hydrodynamical simulations suggest an outflow model as a potential alternative to the high-spin model, proposing that the internal feedback from supernovae is relatively weak compared to typical dwarf galaxies. Under this model, feedback-driven outflows weaken central gravitational potentials, leading to outward migration of stars and dark matter, ultimately forming an extended stellar disk \citep[e.g.,][]{DiCintio17,Chan18,Cardona-Barrero20}.

In addition to these prominent formation models, a study by \cite{Pina20} suggests an alternative scenario where UDGs exhibit comparable halo spin but a higher angular momentum conversion factor compared to typical dwarf galaxies. Furthermore, \cite{Wrigts21} propose that UDGs may arise from the merger of dwarf galaxies, with a hypothesis of transient amplification of descendant halos' spin during merging events to replicate UDGs. 

In high-density environments, UDGs may also arise from typical dwarf galaxies through tidal interactions with massive galaxies \citep{Carleton19,Jiang19}, or via ram-pressure stripping \citep{Grishin21}. These formation scenarios of UDGs in high-density regions are further supported by spectroscopic investigations of UDG kinematics, revealing minimal rotational motion in UDGs \citep[e.g.,][]{Chilingarian19,vanDokkum19b}. 

Furthermore, \cite{vanDokkum15} propose a failed L$^{\star}$ galaxy (FLG) formation model for UDGs, suggesting that UDGs inhabit L$^{\star}$-type massive dark matter halos but exhibit diminished stellar mass due to environmental influences during the early epochs of the universe \citep{Yozin15}.

Indeed, UDGs located in high-density regions, such as galaxy clusters, were previously thought to be enclosed by substantial dark matter halos to shield them from tidal disruption, potentially representing FLGs. Notably, the largest UDG in the Coma cluster, DF44, was initially inferred to harbor a dark matter halo mass of $10^{12}\ \rm{M_{\odot}}$ \citep{vanDokkum16}, based on the dynamics of member globular clusters (GCs). However, subsequent assessments of halo mass using stellar disk kinematics have cast doubt on this assumption of a massive halo \citep{vanDokkum19b}.

Further statistical analyses of UDG samples, utilizing the number/mass of GCs and weak lensing, suggest that the majority of UDGs in galaxy clusters possess masses akin to typical dwarf galaxies ($\sim 10^{11}\ \rm{M_{\odot}}$), contrasting with the mass of the Milky Way \citep{Sifon18,Peng16}. Additionally, spectroscopic observations of GCs in two UDGs within the NGC1052 galaxy group reveal the absence of dark matter components within these UDGs \citep{vanDokkum18,vanDokkum19}. Current understanding leans towards UDGs residing in low-mass halos rather than massive ones, with some studies proposing that UDGs are an extension of dEs/dSphs rather than a distinct class of galaxies \citep[e.g.,][]{Zoller23,Venhola17,Conselice18}.

If UDGs indeed exist in low-mass halos, they may exhibit a (thick) disk-like morphology with S\'ersic indices of $n\lesssim 1$ \citep{Leisman17,Rong20b,Rong24b,vanderBurg16}, with or without a central NSC. However, the presence of a bulge-like structure with significant size in their central regions, a feature commonly observed in massive galaxies \citep{Kormendy10,Kormendy12}, is unexpected in UDGs. In this study, we present, for the first time, a serendipitous sample of UDGs showcasing central pseudo-bulges. This novel finding may add complexity to the understanding of UDG halo masses and formation mechanisms. In section~\ref{sec:2}, we introduce observational data and highlight a cohort of UDGs displaying pseudo-bulge structures at their cores, surrounded by extremely faint and expansive stellar disks. Section~\ref{sec:3} delves into the potential formation mechanism of this specific UDG subgroup. Our conclusions are encapsulated in section~\ref{sec:4}.  Throughout this paper, we employ ``$\log$'' to represent ``$\log_{10}$''.

\section{HI-bearing UDGs with pseudo-bulges}\label{sec:2}

\subsection{Sample and optical photometry}\label{sec:2.1}

Our sample is derived from the Arecibo Legacy Fast Alfa Survey \citep[ALFALFA;][]{Giovanelli05,Haynes18}, a comprehensive extragalactic HI survey spanning approximately 6,600 deg$^2$ in high Galactic latitudes. These HI sources have been cross-referenced with optical data from the Sloan Digital Sky Survey \citep[SDSS DR12;][]{Alam15}. Exploiting the increased depth of the recent DESI Legacy Imaging Surveys \citep{Dey19}, we conducted a comparative analysis of the SDSS and DESI images of these HI samples, resulting in the detection of a dwarf galaxy, AGC238976. While AGC238976 appears as a point source in the SDSS images, the DESI images reveal a subtle stellar disk at the galaxy's periphery.

\begin{figure*}
\centering
\includegraphics[width=\textwidth]{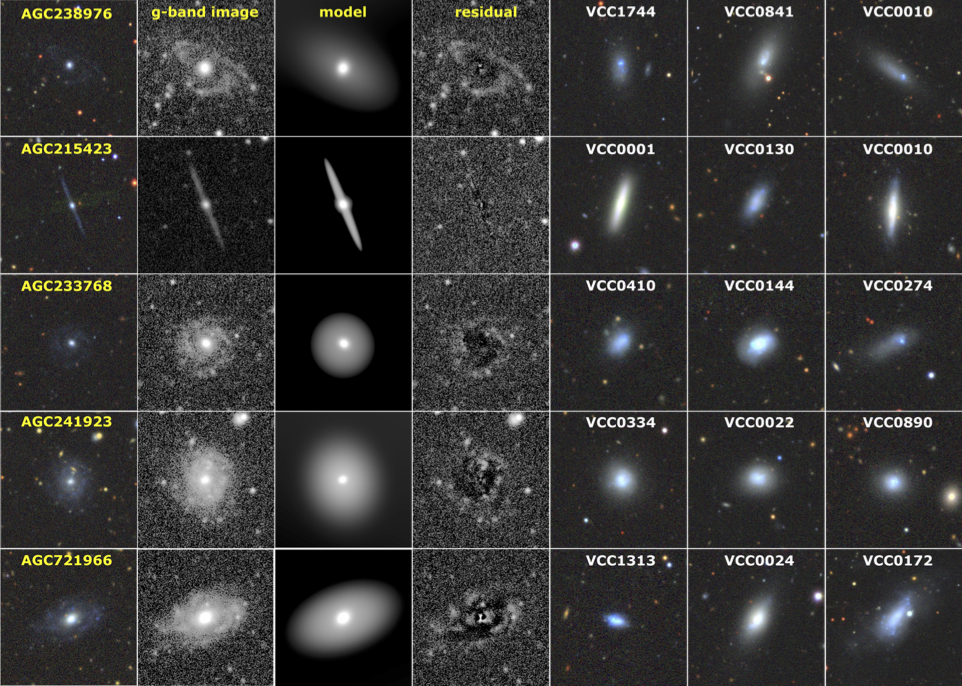}
\caption{The left four columns of panels depict images of the five chosen UDGs with bulges. The panels, from left to right, display the colored DECaLS images, $g$-band images, double-S\'ersic profile fitting outcomes, and residuals. The right three columns of panels showcase DESI images of 15 BCD examples from  \cite{Meyer14} for comparison.}
\label{image}
\end{figure*}

Illustrated in Fig.~\ref{image}, AGC238976 displays a prominent central nuclei-like structure, potentially indicative of a small bulge or NSC, alongside a faint outer disk component. The disk component is particularly discernible in the $g$- and $r$-bands, but less pronounced in the $z$-band. Notably, spiral arm-like structures are also discernible within the disk component.

Following a methodology akin to that outlined in \cite{Zhang}, we employ a `double-S\'ersic' model combined with a `sky' model to fit the $g$, $r$, and $z$-band DESI DR9 images of  AGC238976 using the \textsc{Galfit} software. The two S\'ersic models represent the central nuclei-like structure and the outer disk, respectively. Throughout the fitting procedure, the point spread function (PSF) image for each band is utilized to deconvolve the corresponding galactic image, facilitating the precise determination of the effective radius of AGC238976's central nuclei-like structure. The optimal parameters of the double-S\'ersic model are presented in Table~\ref{fit}. 

\begin{figure*}
\centering
\includegraphics[width=\textwidth]{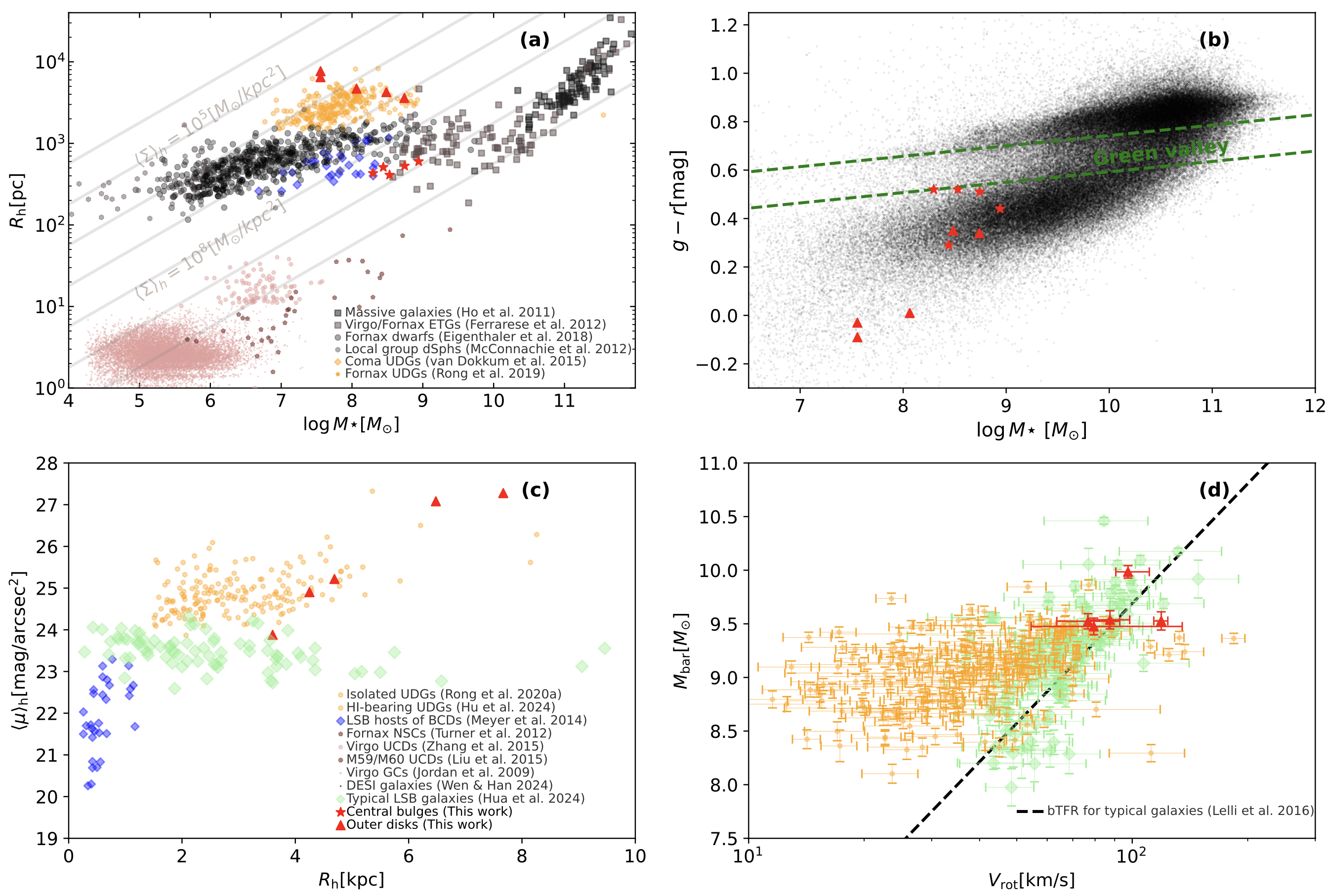}
\caption{a) Stellar mass-size relationship for galaxies and globular clusters, including UDGs. b) Color-stellar mass diagram for DESI galaxies with radial velocities $v<18,000$~km/s, based on data from \cite{Wen24}. {{c) Comparison of effective radius $R_{\rm{h,d}}$ versus $r$-band mean surface brightness $\langle\mu\rangle_{\rm{h,d}}$ for the outer disk components of the five chosen UDGs (red), the LSB hosts of typical BCDs from \cite[blue]{Meyer14}, the stellar disks of typical UDGs from \cite[black]{Hu23}, and the stellar disks of typical LSB galaxies from \cite[light-green]{Hua24}. d) Comparison of baryonic Tully-Fisher relation (bTFR) for the five chosen UDGs hosting pseudo-bulges (red), typical UDGs \citep[orange]{Hu23}, as well as typical LSB galaxies \citep[light-green]{Hua24}. The black dashed line depicts the best-fit bTFR for typical galaxies by \cite{Lelli16}.}} In these panels, the central bulge and outer disk components of the five selected UDGs are represented by red stars and triangles, respectively. The stellar masses of the bulge and disk components of the chosen UDGs are derived from mass-to-light ratios of \cite{Bell03}, with $\log (M_{\star}/L_{\star})_g=1.519\times (g-r)-0.499-0.093$, adjusting the masses by 0.093~dex to transition from a ``diet'' Salpeter to Chabrier initial mass function \citep{Gallazzi08}.}
\label{compare}
\end{figure*}

\cite{Zhang} have demonstrated that a point-like structure with an effective radius exceeding 1/3 of the Full Width at Half Maximum (FWHM) of the PSF can be accurately modeled through fitting. In our analysis of AGC238976, we observe effective radii  $R_{\rm{h}}$ for the central nuclei-like structure of 0.96, 1.12, and 1.03~arcsec in the $g$, $r$, and $z$-bands, respectively. These values surpass the FWHM of the PSF, which are approximately 1.29, 0.94, and 0.89~arcsec for $g$, $r$, and $z$-bands, respectively, enabling precise measurements. This also suggests the presence of a central bulge in AGC238976 with $R_{\rm{h}}\sim 500$~pc, rather than a NSC or GC with $R_{\rm{h}}\sim 3\--100$~pc, as depicted in panel~a of Fig.~\ref{compare}. Considering AGC238976's classification as a dwarf galaxy with a stellar mass of $10^{8.38}\ M_{\odot}$ \citep{Durbala20}, identification of a bulge at its core is particularly intriguing.

The outer disk of AGC238976 exhibits a substantial effective radius of around 7~kpc and an extremely low surface brightness, characterized by an $r$-band mean surface brightness within the effective radius of $\langle \mu \rangle_{\rm{h,d}} \sim 27\ \rm mag/arcsec^2$, meeting the selection criteria for UDGs \citep{vanDokkum15,vanderBurg16}. Therefore, we propose that AGC238976 could be classified as a UDG with a central bulge.

Our subsequent investigation entails a search within the ALFALFA catalog for dwarf galaxies resembling AGC238976, characterized by distinct central nuclei-like sources suggestive of NSCs or minor bulge structures, surrounded by faint stellar disk components. Employing a fitting strategy akin to that of AGC238976, we meticulously analyze their optical images using a `double-S\'ersic' profile. The criteria for identifying analogous galaxies are as follows:

(1) The galaxy's stellar mass should not exceed $10^9\ M_{\odot}$. Previous studies by \cite{Durbala20} involved stellar mass estimations of ALFALFA galaxies through three methods: UV-optical-infrared SED fitting, SDSS $g-i$ color, and $W_2$ magnitude. Preference is given to the stellar mass derived from SED fitting. In cases where UV or infrared data are unavailable for SED fitting, leading to an inability to estimate the stellar mass, the stellar mass is determined based on the $g-i$ color. Any discrepancies in stellar mass obtained from these methods are considered insignificant.

(2) The effective radius of the central nuclei-like structure should range from 1/3 FWHM of {{the}} PSF to 600~pc in the $g$, $r$, and $z$-bands of the DESI survey. This range aligns with the compact systems selection criterion detailed in \cite{Zhang}. The publicly available distance provided by the ALFALFA team \citep{Haynes18} is utilized to calculate the effective radius or each HI-bearing galaxy.

(3) The effective radius $R_{\rm{h,d}}$ of the disk component should exceed 1.5~kpc, with the mean surface brightness within $R_{\rm{h,d}}$ in the $r$-band, $\langle\mu\rangle_{\rm{h,d}}$, not exceeding approximately 24~mag/arcsec$^2$. This criterion is instrumental in identifying UDGs \citep{vanDokkum15,vanderBurg16}.

Only 5 UDGs with central nuclei-like components have been discovered {{from the parent sample with totally 8,600 member dwarf galaxies}}, with AGC238976 being one of them. The photometric results of these five galaxies are elaborated in Table~\ref{fit}, and their DESI images are presented in Fig.~\ref{image}. The effective radii of the central nuclei-like components in all five UDGs significantly surpass the typical range of NSCs, UCDs, and GCs, which typically have effective radii ranging from approximately 3~pc and 100~pc \citep{Walcher06,Georgiev16,Spengler17,Neumayer20,Hilker99,Drinkwater00}, as illustrated in panel~a of Fig.~\ref{compare}. Therefore, these nuclei-like components are identified as bulges within UDGs rather than NSCs.

Moreover, using the methodology outlined by \cite{Rong24a}, we have caculated the distances of these five UDGs from {{their nearest}} galaxy clusters and groups. Importantly, all these UDGs are located beyond three times the virial radii of the galaxy clusters/groups, indicating their isolation from the gravitational influence of these large-scale structures or other galaxies. Hence, the distance measurements of these galaxies are minimally affected by their peculiar velocities. Consequently, these UDGs are also unlikely to be massive late-type galaxies situated at greater distances, as any misinterpretation of their distances could lead to their misclassification as UDGs. In summary, a cohort of UDGs with central bulges has indeed been identified within the ALFALFA sample.

Additionally, it is noteworthy that the S\'ersic indices of the bulge components are below 2.5, indicating a potential classification as pseudo-bulges \citep{Kormendy04}. The S\'ersic indices of the outer stellar disks are below 1, akin to typical isolated UDGs documented by \cite{Leisman17} and \cite{Rong20}.

 On the color versus stellar mass diagram, the outer stellar disks occupy the bluest sector of the `blue cloud', as illustrated in panel~b of Fig.~\ref{compare}, indicating young stellar populations within the stellar disks. Furthermore, these outer disks exhibit spiral-like features, suggesting recent gas accretion and star formation within the stellar disks. The central bulges display relatively redder colors compared to the blue outer disks, with the exception of AGC233768 (where the colors of the bulge and disk are comparable, with $g-r\sim 0.3$~mag), indicating a possible recent quenching or rejuvenation process in the bulges. In essence, these pseudo-bulges are anticipated to have formed earlier than the outer disk components.

\begin{table*}  \centering 
{\fontsize{7}{6}\selectfont
\begin{tabular}{@{}ccccccc|c|cccccccc@{}}
\hline
\hline
\hline
AGC & Dist. & $\log M_{\rm st}$ & $\log M_{\rm HI}$ & $W_{50}$ & Inc. & $V_{\rm{rot}}$ & Band & Comp. & Mag. & $R_{\rm{h}}$ & $\mu_{0}$ & $n$ & $b/a$ & P.A. & $\chi^2/dof$\\
& [Mpc] & [$\log M_{\odot}$] & [$\log M_{\odot}$] & [km/s] & [deg] & [km/s] &  &  & [mag] & [kpc] & [mag/arcsec$^2$] &  &  & [deg] &\\
\hline



& & & & & & & $g$ & disk & 19.31 & 6.75 & 26.15 & 0.78 & 0.55 & 57.6 & 0.587\\
& & &  & & & & & bulge & 18.62 & 0.43 & 19.00 & 1.07 & 1.0 & -6.6 &\\
 \cline{8-16}
238976 & 105.4$\pm 2.3$ & 8.38$\pm 0.03$ & 9.85$\pm 0.05$  & 183$\pm 3$ & $69.9_{-12.8}^{+13.7}$ & $97.5_{-6.9}^{+13.3}$ & $r$ & disk & 19.40 & 7.67 & 26.95 & 0.48 & 0.37 & 54.0 & 0.710\\
& & &  & & & & & bulge & 18.11 & 0.53 & 18.34 & 1.41 & 0.98 & 18.5 &\\
 \cline{8-16}
& & &  & & & & $z$ & disk & 20.17 & 8.66 & 28.31 & 0.06 & 0.15 & 51.1 & 0.792\\
& & &  & & & & & bulge & 17.90 & 0.56 & 18.69 & 1.16 & 0.96 & -19.9 &\\

\hline
\hline

& & &  & & & & $g$ & disk & 19.24 & 6.69 & 27.12 & 0.10 & 0.07 & 21.6 & 0.655\\
& & &  & & & & & bulge & 18.89 & 0.34 & 18.92 & 1.15 & 0.84 & 36.3 &\\
 \cline{8-16}
215423 & 91.8$\pm 2.3$ & 8.03$\pm 0.06$ & 9.40$\pm 0.06$ & 175$\pm 22$ & 90.0 & $87.5_{-11.0}^{+11.0}$ & $r$ & disk & 19.27 & 6.48 & 27.08 & 0.06 & 0.06 & 21.6 & 0.730\\
& & &  & & & & & bulge & 18.37 & 0.41 & 16.36 & 2.45 & 0.88 & 19.4 &\\
 \cline{8-16}
& & &  & & & & $z$ & disk & 19.13 & 5.95 & 27.60 & 0.15 & 0.07 & 22.7 & 0.806\\
& & &  & & & & & bulge & 18.33 & 0.42 & 18.96 & 1.07 & 0.68 & 19.9 &\\

\hline
\hline

& & &  & & & & $g$ & disk & 18.85 & 4.29 & 25.10 & 0.35 & 0.97 & 24.4 & 0.639\\
& & &  & & & & & bulge & 18.73 & 0.52 & 18.30 & 1.64 & 0.84 & 43.8 &\\
 \cline{8-16}
233768 & 115.2$\pm 2.4$ & 8.44$\pm 0.06$ & 9.31$\pm 0.05$ & 60$\pm 6$ & $22.2_{-8.0}^{+7.5}$ & $79.4_{-24.9}^{+55.6}$ & $r$ & disk & 18.50 & 4.25 & 24.18 & 0.75 & 0.94 & 6.80 & 0.714\\
& & &  & & & & & bulge & 18.44 & 0.51 & 18.43 & 1.39 & 0.78 & 55.22 &\\
 \cline{8-16}
& & &  & & & & $z$ & disk & 17.87 & 8.08 & 21.73 & 2.51 & 0.87 & -40.9 & 0.799\\
& & &  & & & & & bulge & 18.20 & 0.55 & 17.98 & 1.59 & 0.72 & 52.32 &\\

\hline
\hline
& & &  & & & & $g$ & disk & 17.56 & 3.73 & 23.80 & 0.60 & 0.84 & 10.7 & 0.518\\
& & &  & & & & & bulge & 19.35 & 0.43 & 20.86 & 0.63 & 0.83 & -59.8 &\\
 \cline{8-16}
241923 & 86.7$\pm 2.4$ & 8.75$\pm 0.05$ & 9.32$\pm 0.05$ & 86$\pm 6$ & $34.1_{-4.4}^{+4.9}$ & $76.7_{-13.1}^{+16.1}$ & $r$ & disk & 17.22 & 3.60 & 23.33 & 0.64 & 0.87 & 7.4 & 0.556\\
& & &  & & & & & bulge & 18.83 & 0.43 & 20.13 & 0.77 & 0.98 & -66.5 &\\
 \cline{8-16}
& & &  & & & & $z$ & disk & 17.07 & 3.67 & 23.22 & 0.64 & 0.78 & 1.7 & 0.704\\
& & &  & & & & & bulge & 18.61 & 0.48 & 20.13 & 0.78 & 0.76 & -77.3 &\\

\hline
\hline
& & &  & & & & $g$ & disk & 17.79 & 4.59 & 24.98 & 0.37 & 0.58 & -70.0 & 0.494\\
& & &  & & & & & bulge & 17.23 & 0.40 & 15.89 & 2.25 & 0.74 & -43.0 &\\
 \cline{8-16}
721966 & 78.9$\pm 2.3$ & 8.59$\pm 0.06$ & 9.35$\pm 0.06$ & 196$\pm 5$ & $55.4_{-1.2}^{+1.6}$ & $119.0_{-5.2}^{+4.8}$ & $r$ & disk & 17.78 & 4.69 & 25.14 & 0.26 & 0.59 & -70.2 & 0.595\\
& & &  & & & & & bulge & 16.79 & 0.60 & 16.62 & 2.10 & 0.72 & -54.3 &\\
 \cline{8-16}
& & &  & & & & $z$ & disk & 17.62 & 4.45 & 24.63 & 0.47 & 0.55 & -76.1 & 0.724\\
& & &  & & & & & bulge & 16.62 & 0.62 & 16.92& 1.89 & 0.69 & -53.0 &\\
\hline
\hline
\hline
\end{tabular}
\caption{Properties of five UDGs with pseudo-bulges.
Col. (1): ALFALFA name; Col. (2): distance to us; Col. (3): stellar mass \citep{Durbala20}; Col. (4): HI mass; Col. (5): 50\% peak width of HI line; Col. (6): inclination; Col. (7): rotation velocity; Col. (8): photometry bands $g$, $r$, and $z$; Col. (9): bulge or disk components; Col. (10): apparent magnitude corrected for the Galactic extinction; Col. (11): effective radius; {{Col. (12): central surface brightness;}} Col. (13): S\'ersic index; Col. (14): axis ratio; Col. (15): position angle; (15): normalized $\chi^2$.}
}
\label{fit}
\end{table*}

\subsection{HI detection}\label{sec:2.2}

\begin{figure}
\centering
\includegraphics[width=\columnwidth]{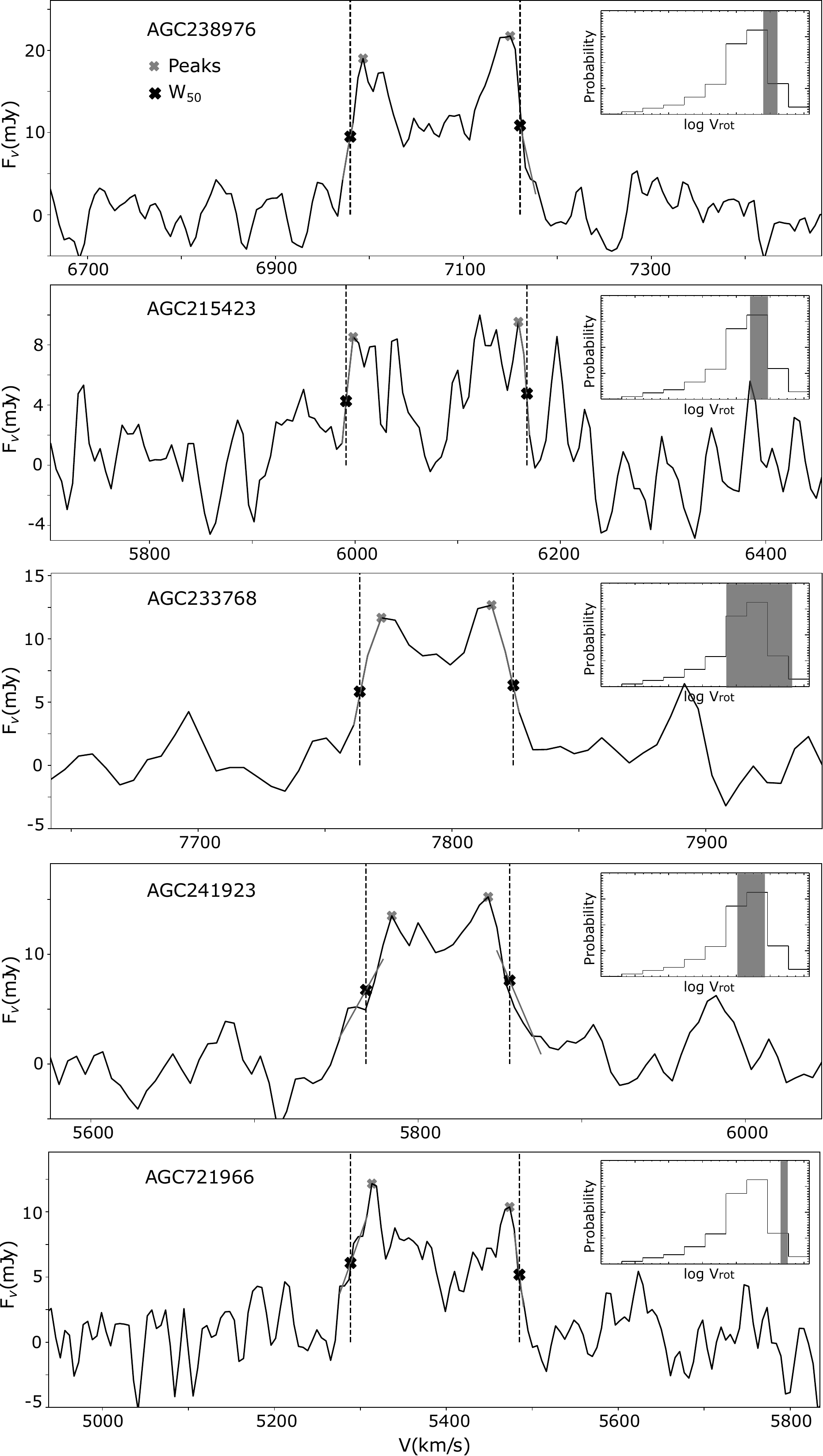}
\caption{HI spectra of the 5 UDGs with bulges from ALFALFA are presented. Each panel displays the positions used for calculating $W_{50}$ for the respective galaxy. The inset in each panel compares the rotation velocity (with the gray region indicating the $1\sigma$ uncertainty range, {{estimated from the errors of $W_{50}$ and inclination.}}) of the galaxy with the rotation velocity distribution (shown as a black histogram) of all HI-bearing dwarf galaxies with $8<\log M_{\rm{st}}/M_{\odot}<9$ in ALFALFA.} 
\label{HI}
\end{figure}

In the left panels of Fig.~\ref{HI}, we present the ALFALFA HI spectra for the five UDGs, with $W_{50}$ representing the 50\% peak widths of the HI lines. In galaxies exhibiting substantial rotation and a relatively large inclination (i.e., small apparent axis ratio $b/a\ll 1$), the HI spectrum is expected to display a double-horned profile with sharp edges on both sides of the HI line \citep{El-Badry18}. In such cases, $W_{50}$ is primarily influenced by the rotation velocity. Conversely, for galaxies with a large apparent axis ratio and low inclination, the HI spectrum should exhibit a single-horned profile, and $W_{50}$ should encapsulate information about both the rotation velocity and velocity dispersion, with a potential dominance of velocity dispersion. However, our five UDGs consistently display double-horned HI spectra, despite the small (optical) inclinations with $b/a\sim 0.9$ observed in the disk components of AGC233768 and AGC241923. This implies that the rotation of gas, and consequently the disk components, in these UDGs are notably significant.

For our five UDGs, we estimate their rotation velocities $V_{\rm{rot}}$ as $V_{\rm{rot}}\simeq W_{50}/2/\sin\phi$, where $\phi$ represents the HI inclination, and $W_{50}$ is corrected for instrumental broadening \citep{Haynes18}. In the absence of resolved HI data, we estimate the HI disk inclination $\phi$ using the average $b/a$ ratio from the $g$, $r$, and $z$-bands, with $\sin\phi=\sqrt{(1-(b/a)^2)/(1-q_0^2)}$ (if $b/a\leq q_0$, $\phi=90^{\circ}$), where $q_0$ denotes the intrinsic thickness of a galaxy. 

Coincidentally, among the UDGs with pseudo-bulges, one galaxy, AGC215423, is observed in an edge-on orientation (Fig.~\ref{image}), showcasing an exceptionally slim stellar disk with $q_0\lesssim 0.1$. To maintain uniformity, we adopt $q_0\simeq 0.1$ for all five UDGs to determine their rotation velocities, which are summarized in Table~\ref{fit}. 

We further compare the rotation velocities of the five UDGs with those of other HI-bearing dwarf galaxies of similar stellar masses in ALFALFA. Employing an {{average intrinsic thickness of $q_0\simeq 0.2${\footnote{{{Based on the findings of \citep{Rong24b}, the intrinsic thickness of young stellar population components, which reflect the gas distributions in dwarf galaxies, may be thicker, with $q_0\sim 0.4$. Using $q_0\sim 0.4$ or $q_0\sim 0.2$ for the parent sample would not alter our conclusion.}}}} for the HI-bearing galaxies \citep{Tully09,Giovanelli97,Li21}, we estimate their rotation velocities.}}  As illustrated in the insets of  Fig.~\ref{HI}, the rotation velocities of the five UDGs are relatively high compared to the broader sample of HI-bearing dwarfs within the stellar mass range of $8<\log M_{\rm{st}}/M_{\odot}<9$, potentially indicating the presence of massive halos or high spins in these UDGs \citep{Mo98,Rong24a}. 

\section{Discussion}\label{sec:3}

\subsection{Are these galaxies exceptional?}\label{sec:3.1}

Dwarf galaxies with bulges are not rare. One notable subset is the blue compact dwarf galaxies (BCDs), characterized by high interstellar matter density, vigorous star formation activity, small effective radii \citep{Janowiecki14,Meyer14} and low S\'ersic indices $n<2.5$ \citep{Lian15,Zhang}. Observations have revealed that a significant proportion of BCDs feature stellar disks, often referred to as ``hosts'', with low surface brightness \citep[LSB;][]{Meyer14} or the accretion of gas leading to the formation of HI disks \citep[e.g.,][]{Tang22}. Hence, from a structural standpoint, both BCDs and the five selected UDGs exhibit bulge-like structures and outer faint disks, showcasing similarities.

Nevertheless, a comparison between the outer disk components of our five sources and the LSB hosts of BCDs, as depicted in panels c and a of Fig.~\ref{compare}, reveals that the outer disk components of our UDGs are considerably larger and fainter than those of the LSB hosts of BCDs, bearing a closer resemblance to UDGs. Furthermore, a visual inspection of the morphologies presented in Fig.\ref{image} highlights distinct differences between BCDs and our five UDGs.

Regarding UDGs, while some UDGs have been identified to host NSCs in their centers, the presence of pseudo-bulges in the central regions of UDGs remains unreported. {{However, there is growing evidence suggesting that distinct populations of UDGs with likely different formation pathways exist  \citep{Buzzo24,Jones22}}}. Hence, the characteristics exhibited by the five UDGs we have uncovered appear to be distinctly unique, whether viewed through the lens of BCDs or UDGs.

\subsection{Formation of these UDGs with pseudo-bulges}\label{sec:3.2}

One intriguing question pertains to the evolutionary mechanism behind the transition of these sub-luminous galaxies from a bulge-shaped structure in their early formation to a disk structure in more recent formations. Empirical galaxy formation models suggest that the likelihood of a galaxy adopting a disk structure is higher if its halo possesses a higher spin, whereas a halo with lower spin is more inclined to form a dense, compact structure resembling a bulge shape \citep[e.g.,][]{Mo98,vandenBosch98,Diemand05,Desmond17,Liao19}. Hence, the formation of bulges characterized by relatively older stellar populations is likely to occur within dark matter halos exhibiting low spin values, whereas the emergence of outer disks, distinguished by high rotation velocities, is expected within halos characterized by high spin values, thus presenting an inherent paradox.

{{One plausible explanation for this paradox}} suggests that BCD-like pseudo-bulges may have originated from galaxy-galaxy mergers \citep{Bekki08}. The merging of dwarf galaxies can induce central starbursts and give rise to massive compact cores dominated by young stellar populations. Meanwhile, the older stellar components in the precursor dwarf galaxies may transform into diffuse, LSB components post-merger \citep{Bekki08}. However, this scenario suggests that the compact cores are likely to be characterized by younger (bluer) stellar populations and embedded within older (redder) disks with a more diffuse spatial distribution. This model provides a potential explanation for the formation of AGC233768, where {{the outer disk (with $g-r\sim 0.35$~mag) displays a slightly redder hue than the central bulge (with $g-r\sim 0.29$~mag)}}, contrasting with the other four galaxies where the pseudo-bulges are redder, and therefore older, than the outer disks.

Alternatively, we suggest that during the merging phase, the two dark matter halos may have contained minimal or no stars but abundant gas reservoirs. As a result, the merging of the gas components from the two halos could have initiated starbursts, leading to the formation of compact pseudo-bulges without an accompanying old stellar disk. Subsequently, any remaining gas would likely have been expelled from the merged galaxy post-merger due to stellar or AGN feedback mechanisms activated by the merging event.

Concurrently, the significant spin of the halo in the progeny galaxy following the merger, derived from the orbital angular momentum of the merging system, provides an explanation for the heightened rotation velocities observed in our five UDGs. Recently, these bulge-like galaxies have experienced a renewed influx of gas from the neighboring environment \citep{Chandola23}, inheriting the high spin of their halos, reigniting the star formation process, and facilitating the formation of thin stellar disks.

However, this theoretical framework faces a potential limitation, as it presupposes that the precursor halos of the merging galaxies must harbor gas while being devoid of or possessing minimal stars, a condition that remains uncertain in cosmological simulations. {{Additionally, if the five selected UDGs exhibit higher spins, their positions on the baryonic mass versus rotation velocity diagram may significantly differ from the baryonic Tully-Fisher relation (bTFR) observed in typical galaxies \citep{Lelli16}, but align with that of typical UDGs \citep{Rong24a,Hu23}. As shown in panel~d of Fig.~\ref{compare}, due to the small sample size, we are unable to determine whether these five UDGs adhere to the bTFR of typical galaxies or UDGs.}}

{{Moreover, we cannot dismiss the alternative scenario that }}these UDGs might be ensconced within substantial dark matter halos, consequently contributing to their augmented rotation velocities. To ascertain the halo mass of these UDGs, resolved HI observations and gas velocity mappings are imperative. Should these UDGs be confirmed to inhabit massive halos while maintaining stellar masses akin to typical dwarf galaxies, they could potentially represent the ``failed $L^{\star}$ galaxies'' proposed by  \cite{vanDokkum15}. The pseudo-bulges observed in these ultra-diffuse galaxies may have originated from intense early AGN feedback, potentially leading to the cessation of subsequent star formation and the formation of sub-luminous galaxies. Furthermore, the early strong AGN feedback also hints at the possibility of higher central black hole masses in these UDGs compared to conventional dwarf galaxies.  This scenario bears resemblance to the recent discovery by the JWST telescope of high-redshift ``little red dots'' \citep{Matthee24}, characterized by compact nuclei-like structures and a higher prevalence of central black hole masses relative to galaxies of equivalent stellar mass \citep{Durodola24}. In this context, the five UDGs we have identified may represent the descendants of ``little red dots''. 

\section{Conclusion}\label{sec:4}

By analyzing the images of the ALFALFA dwarf galaxies from the DESI Legacy Imaging Survey, we have identified five dwarf galaxies featuring central pseudo-bulges. These galaxies display blue pseudo-bulges with S\'ersic indices $n<2.5$ and effective radii spanning 300-700~pc, along with bluer thin stellar disks exhibiting low surface brightness and expansive effective radii that align with the UDG definition. We compare the properties of their disk components with the LSB hosts of BCDs and stellar disks of typical UDGs, and find the outer disk components resemble UDGs more closely. These galaxies are thus referred as UDGs with pseudo-bulges. 

The rotation velocities of these UDGs, determined through calculations utilizing HI line widths and optical inclinations, are relatively high compared to those of typical dwarf galaxies of equivalent mass, suggesting the presence of high halo spins or substantial halos. 

{{We propose two potential formation models for these UDGs. Firstly, they are likely formed from the merger of dwarf galaxies which lacked old stars in the progenitors, resulting in a central bulge-shaped structure, while concurrently enhancing their halo spin. Subsequent accretion of surrounding gas facilitates the formation of extended stellar disk. }}

{{Secondly, it is also plausible that these UDGs could represent ``failed $L^{\star}$ galaxies'' with massive dark matter halos but significantly reduced star formation efficiencies. It is possible that intense AGN feedback is necessary to account for the low star formation efficiency and bulge-shaped structures of these galaxies, suggesting the presence of overmassive central black holes relative to their host galaxies in the early universe. If this scenario holds true, then the five UDGs may be viewed as descendants of the ``little red dots'' discovered by JWST.}}

\clearpage

\begin{acknowledgments}

We thank Wen, Zhong Lue at NAOC for data. YR acknowledges supports from the NSFC grant 12273037, the CAS Pioneer Hundred Talents Program (Category B), the USTC Research Funds of the Double First-Class Initiative, and the research grants from the China Manned Space Project (the second-stage CSST science projects: ``Investigation of small-scale structures in galaxies and forecasting of observations'' and ``CSST study on specialized galaxies in ultraviolet and multi-band''). HXZ acknowledges support from the NSFC grant 11421303. QG is supported by the National SKA Program of China No. 2022SKA0110201, the CAS Project for Young Scientists in Basic Research Grant No. YSBR-062, and the European Union's Horizon 2020 Research and Innovation Programme under the Marie Sk\l odowska-Curie grant agreement No.~101086388. HYW is supported
by the National Natural Science Foundation of China (NSFC, Nos. 12192224) and CAS Project for Young Scientists in Basic Research, Grant No. YSBR-062.

The Legacy Surveys consist of three individual and complementary projects: the Dark Energy Camera Legacy Survey (DECaLS; Proposal ID \#2014B-0404; PIs: David Schlegel and Arjun Dey), the Beijing-Arizona Sky Survey (BASS; NOAO Prop. ID \#2015A-0801; PIs: Zhou Xu and Xiaohui Fan), and the Mayall z-band Legacy Survey (MzLS; Prop. ID \#2016A-0453; PI: Arjun Dey). DECaLS, BASS and MzLS together include data obtained, respectively, at the Blanco telescope, Cerro Tololo Inter-American Observatory, NSF's NOIRLab; the Bok telescope, Steward Observatory, University of Arizona; and the Mayall telescope, Kitt Peak National Observatory, NOIRLab. Pipeline processing and analyses of the data were supported by NOIRLab and the Lawrence Berkeley National Laboratory (LBNL). The Legacy Surveys project is honored to be permitted to conduct astronomical research on Iolkam Du'ag (Kitt Peak), a mountain with particular significance to the Tohono O'odham Nation.

NOIRLab is operated by the Association of Universities for Research in Astronomy (AURA) under a cooperative agreement with the National Science Foundation. LBNL is managed by the Regents of the University of California under contract to the U.S. Department of Energy.

This project used data obtained with the Dark Energy Camera (DECam), which was constructed by the Dark Energy Survey (DES) collaboration. Funding for the DES Projects has been provided by the U.S. Department of Energy, the U.S. National Science Foundation, the Ministry of Science and Education of Spain, the Science and Technology Facilities Council of the United Kingdom, the Higher Education Funding Council for England, the National Center for Supercomputing Applications at the University of Illinois at Urbana-Champaign, the Kavli Institute of Cosmological Physics at the University of Chicago, Center for Cosmology and Astro-Particle Physics at the Ohio State University, the Mitchell Institute for Fundamental Physics and Astronomy at Texas A\&M University, Financiadora de Estudos e Projetos, Fundacao Carlos Chagas Filho de Amparo, Financiadora de Estudos e Projetos, Fundacao Carlos Chagas Filho de Amparo a Pesquisa do Estado do Rio de Janeiro, Conselho Nacional de Desenvolvimento Cientifico e Tecnologico and the Ministerio da Ciencia, Tecnologia e Inovacao, the Deutsche Forschungsgemeinschaft and the Collaborating Institutions in the Dark Energy Survey. The Collaborating Institutions are Argonne National Laboratory, the University of California at Santa Cruz, the University of Cambridge, Centro de Investigaciones Energeticas, Medioambientales y Tecnologicas-Madrid, the University of Chicago, University College London, the DES-Brazil Consortium, the University of Edinburgh, the Eidgenossische Technische Hochschule (ETH) Zurich, Fermi National Accelerator Laboratory, the University of Illinois at Urbana-Champaign, the Institut de Ciencies de l'Espai (IEEC/CSIC), the Institut de Fisica d'Altes Energies, Lawrence Berkeley National Laboratory, the Ludwig Maximilians Universitat Munchen and the associated Excellence Cluster Universe, the University of Michigan, NSF's NOIRLab, the University of Nottingham, the Ohio State University, the University of Pennsylvania, the University of Portsmouth, SLAC National Accelerator Laboratory, Stanford University, the University of Sussex, and Texas A\&M University.

BASS is a key project of the Telescope Access Program (TAP), which has been funded by the National Astronomical Observatories of China, the Chinese Academy of Sciences (the Strategic Priority Research Program “The Emergence of Cosmological Structures” Grant \# XDB09000000), and the Special Fund for Astronomy from the Ministry of Finance. The BASS is also supported by the External Cooperation Program of Chinese Academy of Sciences (Grant \# 114A11KYSB20160057), and Chinese National Natural Science Foundation (Grant \# 12120101003, \# 11433005).

The Legacy Survey team makes use of data products from the Near-Earth Object Wide-field Infrared Survey Explorer (NEOWISE), which is a project of the Jet Propulsion Laboratory/California Institute of Technology. NEOWISE is funded by the National Aeronautics and Space Administration.

The Legacy Surveys imaging of the DESI footprint is supported by the Director, Office of Science, Office of High Energy Physics of the U.S. Department of Energy under Contract No. DE-AC02-05CH1123, by the National Energy Research Scientific Computing Center, a DOE Office of Science User Facility under the same contract; and by the U.S. National Science Foundation, Division of Astronomical Sciences under Contract No. AST-0950945 to NOAO.

\end{acknowledgments}

%



\bibliography{ms}{}
\bibliographystyle{aasjournal}



\end{document}